\documentclass{PoS}

\leftline{\parbox{20cm}{\large Edinburgh 2016/22, LTH 1115, ADP-16-47/T1003 }} 

\title{Partially conserved axial vector current and applications}

\ShortTitle{Partially conserved axial vector current and applications}

\author{R.~Horsley$^{a}$, S.~Kazmin$^{b}$, Y.~Nakamura$^{c}$, \speaker{H.~Perlt}$^{b}$, P.~E.~L.~Rakow$^{d}$, G.~Schierholz$^{e}$,
A.~Schiller$^{b}$ and
J.~M.~Zanotti$^{f}$
\\
$^{a}$ School of Physics and Astronomy, University of Edinburgh, Edinburgh EH9 3FD, UK
\\
$^{b}$ Institut f\"ur Theoretische Physik, Universit\"at Leipzig, PF 100 920, D-04009 Leipzig, Germany
\\
$^{c}$ RIKEN Advanced Institute for Computational Science, Kobe, Hyogo 650-0047, Japan
\\
$^{d}$ Theoretical Physics Division, Department of Mathematical Sciences, University of Liverpool,
\\
\hspace{2mm} Liverpool L69 3BX, UK
\\
$^{e}$ Deutsches Elektronen-Synchrotron DESY, 22603 Hamburg, Germany
\\
$^{f}$ CSSM, Department of Physics, University of Adelaide, Adelaide SA 5005, Australia
\\
E-mail: \email{perlt@itp.uni-leipzig.de}
}

\abstract{We investigate  implications of the use of the point-split axial
vector current derived from a Wilson like fermionic action. We compute the 
corresponding renormalization factor nonperturbatively for one beta value.
The axial charge gA calculated from this nonlocal current is found to be
nearer to the physical value than computed with the local axial vector current --
computed both on the same lattice with the same action.}

\FullConference{34th annual International Symposium on Lattice Field Theory\\
         24-30 July 2016\\
         University of Southampton, UK}

\begin{document}

\section{Introduction}

It is well known that the lattice local axial vector current $A_\mu^{\rm loc} = \overline{\psi}(x) \gamma_\mu\gamma_5 \psi(x)$
does not satisfy the continuum form of the axial Ward identity
\begin{equation}
 \partial_\mu A_\mu(x) = 2mP(x), \quad  P(x)=\bar{\psi}(x)\gamma_5 \psi(x)\,,
\end{equation}
which is due to lattice artefacts. In most cases, where Wilson like fermions
are used, the corresponding improved renormalized current 
\begin{equation}
 A_\mu^{\rm loc, \overline{MS}}(x) = 
  Z_{A^{\rm loc}}(1+b_A a m)\left[A_\mu^{\rm latt,loc}+a c_A \partial_\mu P^{\rm latt} \right]
\end{equation}
is taken to compute physical quantities, like the nucleon axial charge $g_A$.
It turned out, however, that the resulting $g_A$ value are slightly below the 
experimental number and it requires a large effort to bring the lattice result  into coincidence
with it  \cite{Horsley:2013ayv,Bali:2014nma,Alexandrou:2016xok,Bhattacharya:2016zcn,Collins:2016}.
An alternative possibility is to use the point-split (ps) axial vector current
\begin{equation}
A^{\rm ps}_\mu(x)= \frac{1}{2}\left[\bar \psi_x \gamma_\mu \gamma_5 U_\mu(x) \psi_{x+a \hat \mu}+
   \bar \psi_{x+a \hat \mu} \gamma_\mu \gamma_5 U_\mu^\dagger(x) \psi_{x} \right]\label{Aps}\,,
\end{equation}
which is known to fulfill the corresponding lattice axial Ward identity~\cite{Bochicchio:1985xa}.
In~\cite{Horsley:2015nae} we could show that this identity is fulfilled both perturbatively (in one-loop)
and nonperturbatively for the  SLiNC action~\cite{Cundy:2009yy}.
In this work we present implications of this nonlocal lattice form of the axial current - the 
renormalization and the computation of the axial charge $g_A$.


\section{Renormalization}

It is known that the point-split lattice vector current $V^{\rm ps}_\mu(x)$ as obtained
from the lattice vector Ward identity is conserved. This
leads to the renormalization factor $Z_V^{\rm ps}=1$ whereas the nonconserved local counterpart differs
by a finite number from that. One could speculate what behavior $Z_A^{\rm ps}$ 
computed for (\ref{Aps}) shows. On the one hand $A^{\rm ps}_\mu(x)$ is
also a result of the lattice axial vector Ward identity. On the other
hand there appears an extra term which cannot be absorbed into a 
redefinition of the current. This is due to the fact that
Wilson like actions break chiral symmetry. For the SLiNC action
(where the fermionic part is a stout smeared version of a clover improved Wilson action)
this extra term is
a combination of the standard Wilson term and the clover term.
We perform a nonperturbative calculation 
on a $32^3\times 64$ lattice at $\beta=5.50$ [$a=0.074(2)$ fm].
In order to perform the chiral limit we use five $\kappa$ values along
the flavor symmetric line ($\kappa_l = \kappa_s$)
corresponding to pion masses $M_\pi$ = 470, 440, 400, 340, 290 MeV.

\begin{figure}[!htb]
  \begin{center}
     \includegraphics[scale=1,clip=true]{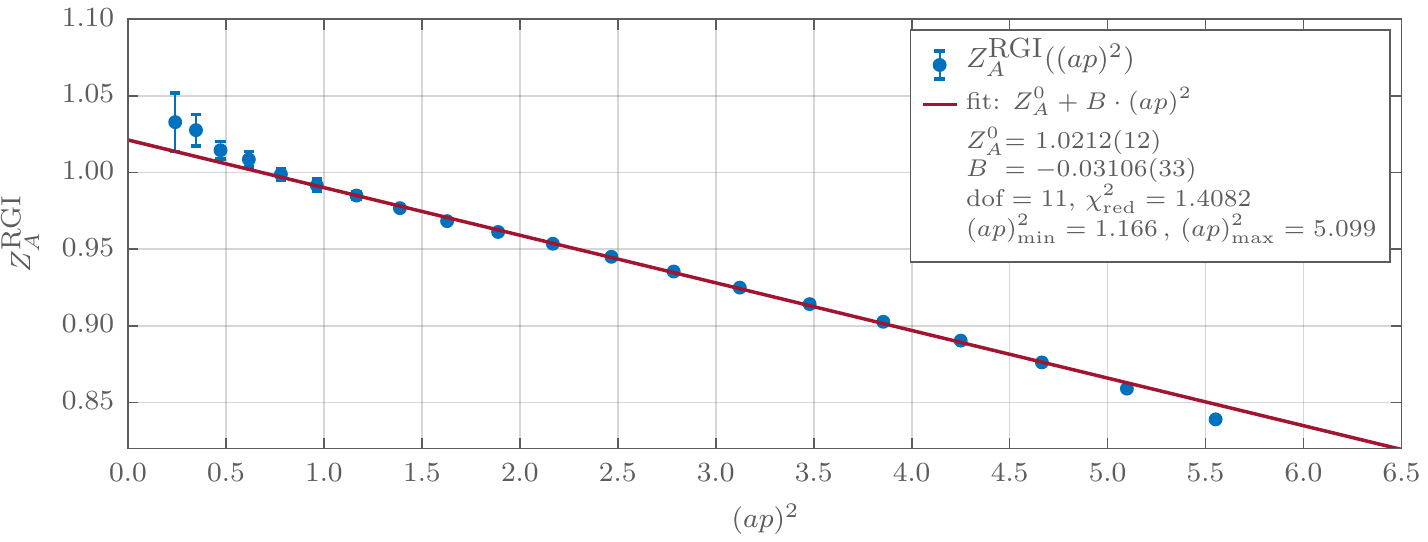}
     \vspace{-3mm}
     \caption{$Z^{\rm RGI}_{A}$ for $\beta=5.5$ as function of $(ap)^2$ for the point-split axial
     current $A=A^{\rm ps}$.}
     \label{fig:ZA}
  \end{center}
\end{figure}

We use the nonperturbative RI$^\prime$-MOM scheme~\cite{Martinelli:1994ty} 
performing a linear chiral extrapolation for each $(ap)^2$ value.
Afterwards we transform  into the RGI and $\overline{\rm MS}$ schemes
which coincide due to the lack of anomalous dimensions.
The result is shown in Fig.~\ref{fig:ZA}.
In the chosen momentum interval $Z_A$ is fitted as a linear function in $(ap)^2$.
Variations of this interval determines the systematic error which dominates
the fit error.
We find $Z_A^{\rm ps} =  1.0212(12)_{\rm fit}(47)_{\rm sys}$ which is very near to one. This 
is consistent with one-loop perturbative results using, however, a different
gauge action~\cite{Horsley:2015nae}.
That result would mean that practically quantities
computed from this point-split axial vector current alone do not need to be
renormalized. It remains to check, however, that this behavior
remains valid for other $\beta$ values (lattice spacings).

\section{Nucleon axial charge $g_A$}

The axial charge $g_A$ is an important quantity to 
understand the spin structure
of the nucleon, but also plays a role in certain astrophysical processes.
It can be measured in the $\beta$ decay of the neutron ($n \rightarrow p + e + \overline{\nu}_e$)
where it determines the angular distribution of the emitted electron.
The current experimental value is given in~\cite{Agashe:2014kda} as
$g_A = 1.2723(23)$.

On the lattice $g_A$ is calculated from the forward matrix element of
the axial vector current $A_{\rm q}^\mu$
\begin{equation}
\langle {\bf p},s | A_{\rm u-d}^\mu |{\bf p},s \rangle = 2\, g_A \,s^\mu \,,
\label{gAlatt}
\end{equation}
where $ |{\bf p},s \rangle$ is a proton state with momentum ${\bf p}$ and spin $s^\mu$ and 
the inserted operator is $A_{\rm u-d}^\mu = A_{\rm u}^\mu - A_{\rm d}^\mu$.
Being a nonsinglet quantity there are no contributions from disconnected
quark lines. The relation (\ref{gAlatt}) makes this observable to a benchmark 
test for lattice calculations. For a review of the current status see~\cite{Collins:2016}.
Despite the progress that has been made in the last years there remain a couple
of challenges to be solved. Among them we mention the extrapolation to the physical point
and the treatment of excited states.

In this work we take the point-split axial vector current (\ref{Aps}) as the 
operator inserted in (\ref{gAlatt}). As lattices we use
\{$32^3\times 64$,  $\beta=5.50$ [$a=0.074(2)$ fm], $M_\pi$ = 470, 360, 310 MeV\}
along the $\bar{m} =$ const. line
and \{$48^3\times 96$,  $\beta=5.80$ [$a=0.059(3)$ fm], $M_\pi$ = 427 MeV\}
at the flavor symmetric point.
$g_A$ has to be estimated from the ratio of the 3-point function to the 2-point function
\begin{equation}
  R(t_i,t_f,\tau) = \frac{G_3(t_i,t_f,\tau)}{G_2(t_i,t_f)} \quad \rightarrow \quad g_A\label{ratio}
\end{equation}
with $(t_f-t_i)$ - the source-sink distance, $\tau$ - the source-operator insertion distance.
It is clear that a meaningful determination of $g_A$ is possible only if the ratio
exhibits a pronounced plateau, ideally independent on $(t_f-t_i)$ and $\tau$.

One of the main challenges of this kind of computations is the handling of excited states.
There are various techniques which are used to take them into account. Among them we
have the summation, the multi-exponential fit and the 
variational methods~\cite{Capitani:2012gj,Owen:2012ts,Yoon:2016dij,Dragos:2016rtx}.

\begin{figure}[!htb]
 \begin{center}
    \includegraphics[scale=1,clip=true]{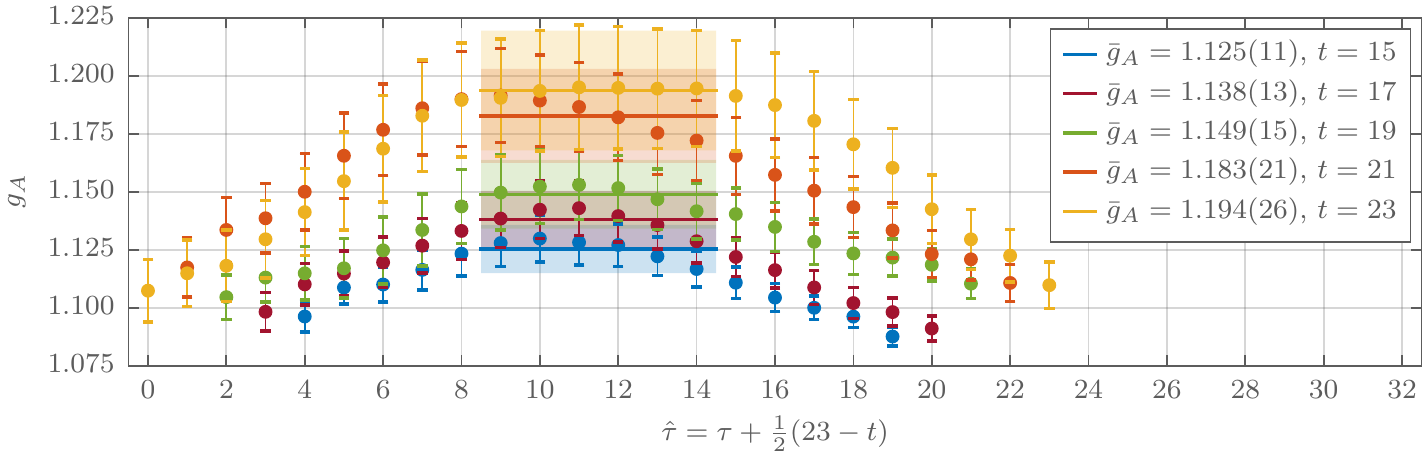}
    \vspace{-3mm}
    \caption{Plateaus of $R(t_i,t_f,\tau)$ for various $t_{sep}=(t_f-t_i)$ as function of operator insertion time $\tau$ for $\beta=5.80$.}
    \label{fig:plateaus}
  \end{center}  
\end{figure}

It turns out that the source-sink distance determines the form and the height of the plateau
of the ratio $R(t_i,t_f,\tau)$ defined in~(\ref{ratio}) (see Fig.~\ref{fig:plateaus}).
This is directly connected to the influence of excited states which diminish the height for smaller separations.
Investigations in~\cite{Dragos:2016rtx} suggest that the variational method shows very stable results.

In our computation, however, we used the 3-exponential fit method which includes the first three
energy levels ($t=t_{\rm sep}$). The fit to the ratio (\ref{ratio}) has the form
\begin{eqnarray}
 F_{g_A}(t,\tau) = g_A^{\rm fit}& & \left[ 1+C_{10}\left(e^{-\tau M_{10}}+
	  e^{-(t-\tau) M_{10}}\right)+C_{11}e^{-t M_{10}}\right.\nonumber\\
 & & +\,C_{20}\left(e^{-\tau M_{20}}+e^{-(t-\tau) M_{20}}\right)+C_{22}e^{-t M_{20}}\nonumber\\
 & & \left.+\,C_{21}\left(e^{-\tau M_{21}}e^{-t M_{10}}\right)+C_{21}\left(e^{-t M_{20}}e^{-\tau M_{21}}\right)\right]\nonumber\\
 & & \times \left[1+D_{10}e^{-t M_{10}}+D_{20}e^{-t M_{20}}\right]^{-1}\,,
 \label{FgA}
 \end{eqnarray}
where $M_{ik}=M_i-M_k$ and the $M_i$ are the masses of the ground state ($i=0$) 
and the next two excited states ($i=1,2$). They can be
determined rather precisely from the corresponding 2-point functions, as shown in Fig.~\ref{fig:corr}.
The fit ({\ref{FgA}) is performed for the parameters $g_A^{\rm fit}, C$ and $D$ over the available
data sets $(t,\tau)$ simultaneously. An example is shown in the right of Fig.~\ref{fig:corr}.
\begin{figure}
  \begin{center}
  \begin{tabular}{cc}
     \includegraphics[scale=1,clip=true]{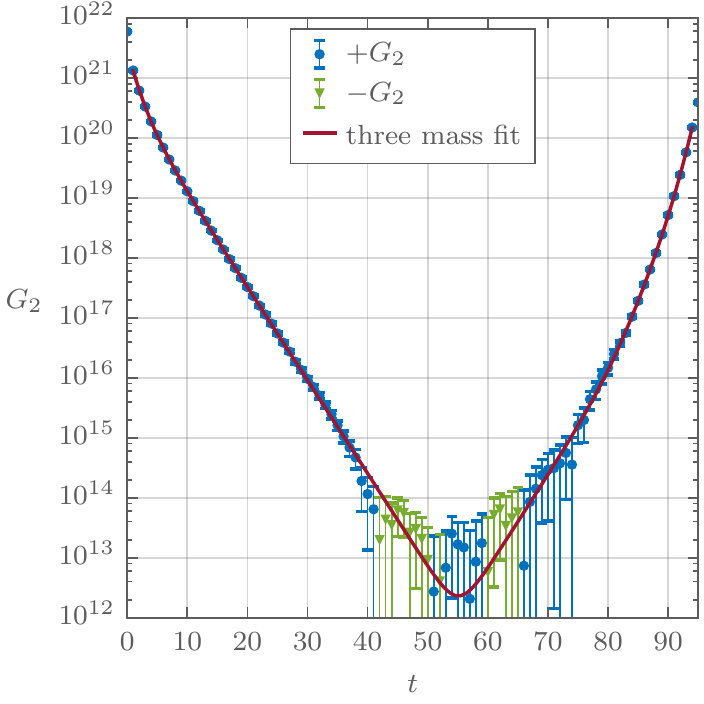}
     &
     \includegraphics[scale=1,clip=true]{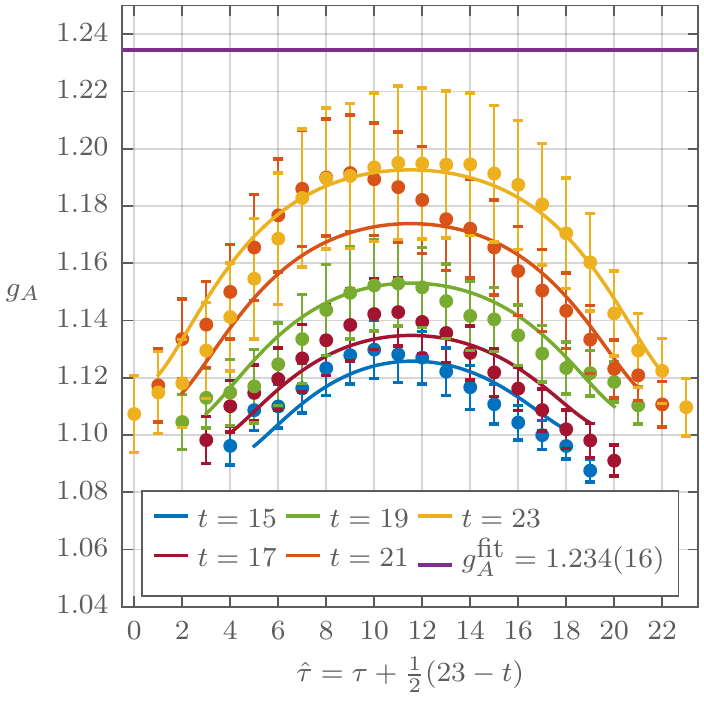}
  \end{tabular}
  \end{center}
  \vspace{-2mm}
  \caption{Left: Three mass fit to the 2-point correlation function for $\beta=5.80$. Right: Result
  of the global fit (\protect\ref{FgA}) for $\beta=5.80$.}
  \label{fig:corr}
\end{figure}
where we fit over the whole data set with all available separations $t_{sep}=15 ... 23$. 

Our final results for $g_A$  at $\beta=5.50$ ($M_\pi$ = 470~MeV) and $\beta=5.80$
are shown in Fig.~\ref{fig:gA}.
\begin{figure}[!htb]
  \begin{center}
     \includegraphics[scale=1,clip=true]{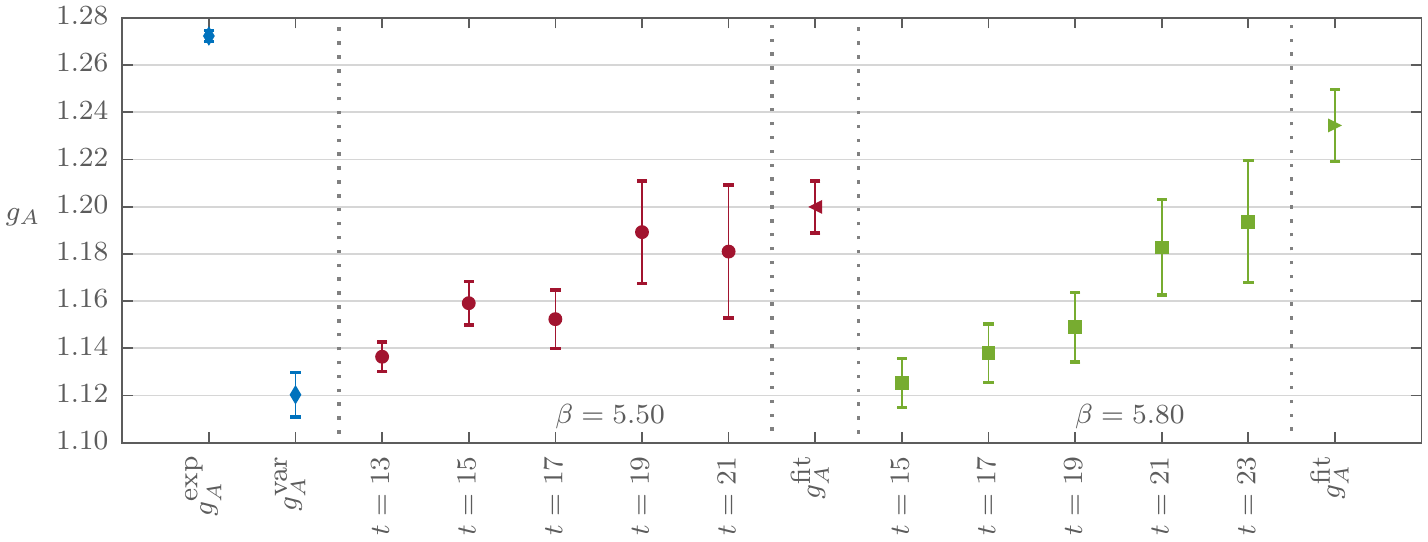}
     \vspace{-3mm}
     \caption{$g_A$ values using different methods at the flavor symmetric points. 
	      $\beta=5.50$ and $M_\pi$ = 470~MeV: $g_A^{\rm var}$: local axial current (variational method),
              point-split axial current (plateau values at $t=t_{sep}=13,15,17,19,21$)
              and 3-exponential fit $g_A^{\rm fit}$;
              $\beta=5.80$: point-split axial current (plateau values at $t=t_{sep}=15,17,19,21,23$)
              and 3-exponential fit $g_A^{\rm fit}$.
              The experimental value is denoted as $g_A^{\rm exp}$.}
     \label{fig:gA}
  \end{center}      
\end{figure}
We compare them for $\beta=5.50$ with the variational method~\cite{Dragos:2016rtx}
which has been obtained for the local axial curent on the same lattice.
It is obvious that the plateau values depend very much on the source-sink separations.
The 3-exponential fit gives a higher value towards the experimental result.
It can be recognized that (using the same lattice parameters) the axial charge 
for the point-split current is
nearer to the experimental value than the value using the local current.
This should be seen also in connection to the corresponding $Z$ factor
($Z_A^{\rm ps} =  1.0212$) which indicates that we are rather close to
the continuum. Furthermore, we did not find a significant dependence on 
the three pion masses at $\beta = 5.50$.
For $\beta=5.80$ the final fit $g_A^{\rm fit}$ is even larger.
This shows a tendency to increase $g_A(\beta)$ with increasing $\beta$ (decreasing 
lattice spacing $a$) which is encouraging. However, for a sound comparison
with experiment it remains to perform 
a careful extrapolation to the physical point.

\section*{Acknowledgements}

The  numerical  configuration  generation  (using  the  BQCD  lattice  QCD  program~\cite{Nakamura:2010qh})
was carried out on the IBM BlueGene/Q
using DIRAC 2 resources (EPCC, Edinburgh, UK) and the BlueGene/P and Q at NIC (J\"ulich, Germany).
The data analysis (using the Chroma software library~\cite{Edwards:2004sx}) was performed
on the SGI ICE 8200 and Cray XC30
at the HLRN (The North-German Supercomputer Alliance).
HP was supported by the DFG grant SCHI 422/10-1. PELR was supported in part by the STFC
under contract ST/G00062X/1 and JMZ by the Australien Research Council 
Grant No. FT100100005 and DP140103067. We thank all funding agencies.

\end{document}